\documentclass[twocolumn,aps,prb,superscriptaddress,longbibliography]{revtex4-2}
\usepackage[utf8]{inputenc}
\usepackage[T1]{fontenc}
\usepackage[english]{babel}
\usepackage[dvipsnames]{xcolor}
\usepackage[colorlinks]{hyperref}
\hypersetup{colorlinks,linkcolor=red,citecolor=blue,urlcolor=blue,final}
\usepackage{lmodern,fancyhdr}
\usepackage{graphicx,float}
\usepackage{enumitem,amsmath,amssymb,amsthm,bm,array,mathdots,mathrsfs,dsfont}

\newcommand{\curlyL}{\mathcal{L}}
\newcommand{\curlyD}{\mathcal{D}}
\newcommand{\curlyS}{\mathcal{S}}
\newcommand{\curlyB}{\mathcal{B}}
\newcommand{\curlyP}{\mathcal{P}}
\newcommand{\curlyQ}{\mathcal{Q}}
\newcommand{\curlyG}{\mathcal{G}}

\newcommand{\intpic}{ ^{ (\text{2}) } }
\newcommand{\rf}{ ^{ (\text{1}) } }
\newcommand{\sts}{_{ \text{ss}}}
\newcommand{\xop}{\hat{x}}
\newcommand{\popr}{\hat{p}}
\newcommand{\aop}{\hat{a}}
\newcommand{\bop}{\hat{b}}

\newcommand{\sop}{\hat{s}}
\newcommand{\sopd}{\hat{s}^\dagger}
\newcommand{\Hop}{\hat{H}}

\newcommand{\Hint}{\hat{V}}
\newcommand{\sgm}[1]{\hat{\sigma}_{#1}}
\newcommand{\sgmt}[1]{\hat{{\tilde\sigma}}_{#1}}

\newcommand{\sgmtv}{\hat{\tilde{\bm{\sigma}}}_i}
\newcommand{\rhoop}{\hat{\rho}}
\newcommand{\rhotot}{\hat\rho_\mathrm{tot}}
\newcommand{\rhoS}{\hat\rho_S}

\newcommand{\rhoBs}{\hat\rho_{B,s}}

\newcommand{\dotrhotot}{\dot{\hat\rho}_\mathrm{tot}}
\newcommand{\dotrhoS}{\dot{\hat\rho}_S}
\newcommand{\wn}{\omega_n}
\newcommand{\wbi}{\omega_{Bi}}
\newcommand{\coupl}[1]{G_{#1}}
\newcommand{\On}{\Omega_n}
\newcommand{\Obi}{\Omega_{Bi}}
\newcommand{\wdr}{\omega_d}
\newcommand{\kdci}{\kappa_{1i}}
\newcommand{\kdpi}{\kappa_{2i}}
\newcommand{\kti}{\kappa_{ti}}
\newcommand{\dn}{\Delta_n}
\newcommand{\dm}{\Delta_m}
\newcommand{\dbi}{\Delta_{Bi}}
\newcommand{\Urf}{\hat{U}_1(t)}
\newcommand{\Urfd}{\hat{U}_1^\dagger(t)}
\newcommand{\Uint}{\hat{U}_\text{2}(t)}
\newcommand{\Uintd}{\hat{U}_\text{2}^\dagger(t)}
\newcommand{\tvar}{\tau}
\newcommand{\tvarp}{\tau'}
\newcommand{\wsingle}{\omega_0}
\newcommand{\Osingle}{\Omega_0}
\newcommand{\coup}{G}
\newcommand{\dsingle}{\Delta_0}
\newcommand{\wb}{\omega_{B}}
\newcommand{\Ob}{\Omega_{B}}
\newcommand{\kdc}{\kappa_1}
\newcommand{\kdp}{\kappa_2}
\newcommand{\kt}{\kappa_t}
\newcommand{\db}{\Delta_{B}}
\newcommand{\Oeff}{\Omega_{0}'}
\newcommand{\diffrate}{\Gamma}
\newcommand{\sqrate}{g}
\newcommand{\dLamb}{\Delta'}
\newcommand{\GammaA}{\Gamma_{\scriptscriptstyle ++}}
\newcommand{\GammaB}{\Gamma_{\scriptscriptstyle +-}}
\newcommand{\GammaC}{\Gamma_{\scriptscriptstyle -+}}
\newcommand{\GammaD}{\Gamma_{\scriptscriptstyle --}}
\newcommand{\gammaB}{\gamma_{\scriptscriptstyle +}}
\newcommand{\gammaC}{\gamma_{\scriptscriptstyle -}}
\newcommand{\GammaBC}{\Gamma_{\scriptscriptstyle \pm\mp}}
\newcommand{\gammaBC}{\gamma_{\scriptscriptstyle \pm}}
\newcommand{\im}{\text{i}}
\newcommand{\levi}[1]{\epsilon_{#1}}
\newcommand{\hc}{{\text H.c.}}
\newcommand{\tr}{\text{tr}}

\newcommand{\fd}[2]{\frac{\text{d} {#1}}{\text{d} {#2}}}
\newcommand{\integral}[3]{ \int_{#1}^{#2} \text{d} {#3} \hspace{1mm} }
\newcommand{\mv}[1]{ \langle #1 \rangle }
\newcommand{\bra}[1]{ \langle #1 \rvert }
\newcommand{\ket}[1]{ \lvert #1 \rangle }
\newcommand{\be}{\begin{equation}}
\newcommand{\ee}{\end{equation}}
\newcommand{\eqnref}[1]{Eq.~\eqref{#1}}
\newcommand{\figref}[1]{Fig.~\ref{#1}}
\newcommand{\secref}[1]{Sec.~\ref{#1}}

\newcommand{\tabref}[1]{Table~\ref{#1}}
\newcommand{\pare}[1]{\left( {#1} \right)}
\newcommand{\spare}[1]{\left[ {#1} \right]}
\newcommand{\cpare}[1]{\left\{ {#1} \right\}}
\newcommand{\pares}[1]{( {#1} )}
\newcommand{\spares}[1]{[ {#1} ]}

\newcommand{\cparelb}{\Big\{}
\newcommand{\cparerb}{\Big\}}

\begin{document}

\title{Effective quantum dynamics induced by a driven two-level-system bath}

\author{Katja Kustura}
\email{katja.kustura@uibk.ac.at}
\affiliation{Institute for Quantum Optics and Quantum Information of the Austrian Academy of Sciences, A-6020 Innsbruck, Austria.}
\affiliation{Institute for Theoretical Physics, University of Innsbruck, A-6020 Innsbruck, Austria.}
\author{Oriol Romero-Isart}
\affiliation{Institute for Quantum Optics and Quantum Information of the Austrian Academy of Sciences, A-6020 Innsbruck, Austria.}
\affiliation{Institute for Theoretical Physics, University of Innsbruck, A-6020 Innsbruck, Austria.}
\author{Carlos Gonzalez-Ballestero}
\affiliation{Institute for Quantum Optics and Quantum Information of the Austrian Academy of Sciences, A-6020 Innsbruck, Austria.}
\affiliation{Institute for Theoretical Physics, University of Innsbruck, A-6020 Innsbruck, Austria.}

\date{\today}

\begin{abstract}
We derive a Born-Markov master equation describing the dissipation induced by a bath of lossy but coherently driven two-level systems (TLS) coupled to a bosonic system via Jaynes-Cummings interaction. We analytically derive all the master equation rates. We characterize these rates for the particular case of a single-mode system coupled to identical TLS. We study the steady state of the system and its exotic properties stemming from the non-thermal stationary state of the driven TLS bath. These properties include dissipative amplification, bath-induced linear instability, and both coherent and dissipative squeezing. The master equation is valid for arbitrarily strong TLS driving, and it can be generalized to include multi-level systems or other system-bath interaction terms, among others. Our work provides a tool to study and characterize TLS-induced decoherence, a key limiting factor in quantum technological devices based on, for instance, superconducting circuits, magnonic systems, or quantum acoustics.
\end{abstract}

\maketitle

\section{Introduction}

Two-level impurities are known to limit the coherence of many systems of both fundamental and technological interest ranging from superconducting qubits~\cite{Burin2015,Klimov2018,Goetz2017,Schloer2019,Burnett2019} and resonators~\cite{Burnett2014,deGraaf2018,Mueller2019}, to electronic transistors~\cite{TenorioPearl2017,Pourkabirian2014}, ion traps~\cite{Brownnutt2015}, solid-state quantum emitters~\cite{Kleinsasser2016,Bauch2018} or acoustic~\cite{Behunin2017,Andersson2021,Scigliuzzo2020}  and magnonic platforms~\cite{Tabuchi2014,Pfirrmann2019,LachanceQuirion2019,Woltersdorf2009,Mihalceanu2018,MaierFlaig2017}. This two-level-system (TLS) induced decoherence becomes especially significant at low temperature where additional bosonic baths are depleted~\cite{Tabuchi2014}. Recent advances in the control of quantum technological platforms have sparked a revived interest in probing and controlling these TLS baths, whose impact can now be accurately measured~\cite{Tabuchi2014} and even suppressed~\cite{Behunin2017,Andersson2021}. An important step toward this goal would be to bridge the gap between the current, platform-dependent, solid-state-based theoretical descriptions of TLS-induced dissipation, such as the standard tunneling model~\cite{Burin2015,vanVleck1964,Mueller2019,Anderson1972,Black1977,Phillips1987}, and the general quantum-mechanical formulation of dissipation widely used in quantum optics. Specifically, the derivation of a master equation describing the decoherence induced by TLS baths is timely. 

Any description of the dissipation discussed above must account for two unconventional aspects: first, the two-level statistics of the bath, whose richer phenomenology in comparison with e.g. bosonic baths~\cite{BreuerPetruccione} makes them the focus of intense research~\cite{Prokofev2000,Lin1985,Caldeira1993}.  Second, and more importantly, the possibility that the TLS forming the bath are subject not only to their intrinsic loss but also to coherent driving, which drives the bath out of thermal equilibrium. As a consequence the standard assumption in open quantum systems, namely a bath in thermal equilibrium, does not apply. Not only can TLS baths be subject to purposeful driving~\cite{Behunin2017,Andersson2021} but, generally, driving the system will unavoidably result in driving of the bath, either through direct interaction with the driving fields or indirectly through the action of the system itself~\cite{Reichert2016,Grabert2018,Grabert2016}. This spurious bath driving can be strong and have important consequences on the system dynamics and response, especially for nanostructures~\cite{Reichert2016,Grabert2018,Grabert2016}. In this paper we derive a Born-Markov master equation describing the open dynamics of a bosonic system in contact with a bath of independent TLS subject to arbitrarily strong coherent driving. We demonstrate the rich system phenomenology arising both from the TLS statistics and from the non-thermal stationary state of the bath, including amplification, steady-state squeezing, and dynamical instabilities for typical parameters in, for instance, microwave cavities. All these features evidence the potential of controllable TLS bath driving as a tool to reduce TLS-induced decoherence and to probe and understand the complex properties of TLS baths~\cite{Behunin2017}. 

\begin{figure}
\centering
\includegraphics[width=\linewidth]{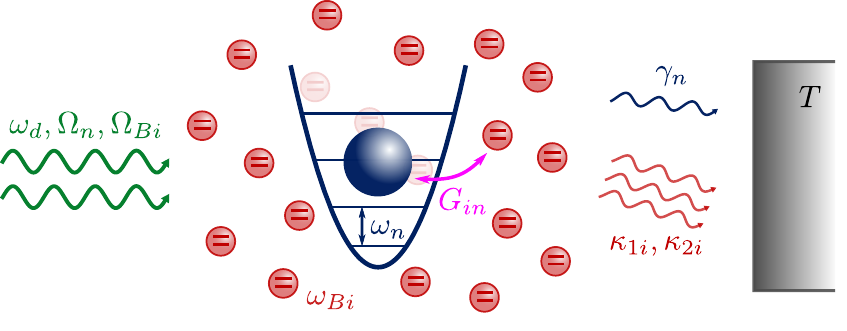}
\caption{Scheme of the model. A set of $N_b$ bosonic modes (here depicted as one for simplicity) with frequencies $\wn$ and decay rates $\gamma_n$ ($n=1,...,N_b)$ is coupled to a bath composed of $N$ two-level systems (TLS) with frequencies $\wbi$ and decay and dephasing rates $\kdci$ and $\kdpi$, respectively ($i=1,...,N$). The $n-$th system mode is coupled to the $i-$th TLS at a rate $\coupl{in}$. Both system and bath are coherently driven at frequency $\wdr$ and with respective amplitudes $\On$ and $\Obi$.}
\label{fig:system}
\end{figure}

Our paper is organized as follows. We start in \secref{sec:model-derivation} by describing the model and outlining the derivation of the master equation for the system. In \secref{sec:case-study}, we focus on the particular case of a single-mode system coupled to a bath of identical TLS. We characterize the behavior of the master equation rates and the system steady state, focusing on the effects arising from the non-thermal state of the TLS bath. The conclusions and outlook are provided in \secref{sec:conclusion}.

\section{The model and the master equation derivation} 
\label{sec:model-derivation}

In this section we describe the derivation of the master equation for the system interacting with a driven TLS bath. We start in \secref{sec:model} by introducing the parameters and the assumptions of our model. We then proceed in \secref{sec:derivation} to give a summary of the master equation derivation under the Born-Markov approximation. 

\subsection{Description of the model}
\label{sec:model}

The system under study, schematically shown in \figref{fig:system}, consists of an ensemble of $N_b$ independent bosonic modes with frequencies $\wn$ ($n=1,2,...,N_b$), each coupled to a finite bath of $N$ non-interacting TLSs with frequencies $\wbi$ ($i=1,2,...,N$). Both the system and the TLS bath are open, i.e. coupled to independent external reservoirs which we model as usual thermal baths at temperature $T$. Additionally, both the system and the TLS bath are subject to coherent driving.  The dynamics of the total system are given by the Liouville--von Neumann equation
\be\label{Liouville-von-Neumann-equation}
    \dotrhotot = (\curlyL_S+\curlyL_B+\curlyL_I)[\rhotot],
\ee
where $\rhotot$ is the total density matrix of the system and the bath, and $\curlyL_S,\curlyL_B$ and $\curlyL_I$ are the Liouvillian superoperators~\cite{BreuerPetruccione} denoting the system, the bath, and the interaction parts, respectively. Let us describe each term separately.

The system Liouvillian $\curlyL_S$ acts only on the system Hilbert space and it is given by the standard optical master equation~\cite{BreuerPetruccione}
\be \label{system-liouvillian}
    \curlyL_S [\rhoop] = -\frac{\im}{\hbar} [\Hop_S, \rhoop] +\curlyS[\rhoop].
\ee
The first term describes the coherent dynamics of the externally driven ensemble of bosonic modes through the Hamiltonian
\be\label{system-Hamiltonian}
\begin{split}
    \Hop_S = \hbar\sum_{n=1}^{N_b} \cparelb 
    \wn\sopd_n\sop_n + \pare{\On \sop_n e^{\im \wdr t} + \hc }
    \cparerb. 
\end{split}
\ee
Here, $\Omega_n \in \mathbb{C}$ is the driving rate for mode $n$, $\omega_d$ the driving frequency, and $\sop_n$ and $\sopd_n$ bosonic annihilation and creation operators obeying the commutation relations $[\sop_m,\sop_n]=[\sopd_m,\sopd_n]=0$ and $[\sop_m,\sopd_n]=\delta_{mn}$. The second term in Eq.~\eqref{system-liouvillian} describes the incoherent dynamics induced by the thermal bath, and it is given by
\begin{multline}\label{system-dissipation}
    \curlyS[\rhoop] = \sum_{n=1}^{N_b} \gamma_n \cparelb 
    \spare{1+\bar{n}(\wn)}  \curlyD_{\sop_n,\sopd_n} [\rhoop]
    \\+ \bar{n}(\wn)           \curlyD_{\sopd_n,\sop_n} [\rhoop] \cparerb,
\end{multline}
where $\gamma_n$ is the decay rate of a mode $n$,  $\curlyD_{\aop,\bop}[\rhoop] = \aop\rhoop\bop-\{\bop\aop, \rhoop\}/2$ is the Lindblad superoperator, and $\bar{n}(\omega)=[\exp(\hbar \omega / k_B T)-1]^{-1}$ is the Bose-Einstein distribution, with $k_B$ the Boltzmann constant.  

The bath contribution $\curlyL_B$ acts only on the Hilbert space of the TLS and it is given by the standard Bloch master equation~\cite{BreuerPetruccione}
\be\label{bath-Liouvillian}
    \curlyL_B [\rhoop] = -\frac{\im}{\hbar} [\Hop_B, \rhoop] +\curlyB[\rhoop].
\ee
The first term describes the coherent dynamics of externally driven TLSs through the Hamiltonian 
\be\label{bath-Hamiltonian}
\begin{split}
    \Hop_B = \frac{\hbar}{2} \sum_{i=1}^N \cparelb
    \wbi\sgm{zi} + \pare{\Obi \sgm{+i} e^{-\im \wdr t} + \hc }
    \cparerb.
\end{split}
\ee
Here, $\Omega_{Bi}\in \mathbb{C}$ is the driving rate for emitter $i$, and we define $\sgm{\pm i}=(\sgm{xi}\pm i \sgm{yi})/2$ in terms of the spin $1/2$ Pauli operators $\{\sgm{xi},\sgm{yi},\sgm{zi}\}$, which obey the commutation relations $[\sgm{\alpha i}, \sgm{\beta j}] = \im \delta_{ij} \levi{\alpha\beta\gamma} \sgm{\gamma i}$ for $\alpha,\beta,\gamma=x,y,z$, with $\levi{\alpha\beta\gamma}$ the Levi-Civita tensor. The second term in Eq.~\eqref{bath-Liouvillian} describes the incoherent TLS dynamics induced by the thermal bath, and it is given by
\be\label{bath-dissipation}
\begin{split}
    \curlyB[\rhoop] = &\sum_{i=1}^N \cparelb
     \kdci\spare{1+\bar{n}(\wbi)}  \curlyD_{\sgm{-i},\sgm{+i}} [\rhoop]\\
    +&\kdci        \bar{n}(\wbi)  \curlyD_{\sgm{+i},\sgm{-i}} [\rhoop]
     +\kdpi                  \curlyD_{\sgm{zi},\sgm{zi}} [\rhoop]\cparerb.
\end{split}
\ee
The terms proportional to $\kdci$ describe decay and absorption, whereas the term proportional to $\kdpi$ describes dephasing. It is useful for the following discussion to define a single decoherence time scale including both processes, namely the TLS transverse decay rate, as
\be\label{transverse-decay-rate}
    \kti = \frac{\kdci}{2}\spare{1+2\bar{n}(\wbi)} + 2\kdpi.
\ee

Finally, the interaction part of the Liouvillian describes the coupling between the system of bosonic modes and the TLS bath. We assume a standard Jaynes-Cummings interaction which, as discussed below, is appropriate to discuss many physical scenarios of interest. Specifically, 
\be \label{interaction-Liouvillian}
\begin{split}
    &\curlyL_I [\rhoop] = -\frac{\im}{\hbar}[\Hint, \rhoop] =
    -\frac{\im}{\hbar}\sum_{n i} \spare{\coupl{in}\sgm{+i}\sop_n + \hc, \rhoop},
\end{split}
\ee
with $\coupl{in}$ the coupling rate between the $i-$th TLS and the $n-$th bosonic system mode. In the above equation and hereafter, we omit for simplicity the upper limit of the sums in $i$ and $n$, namely $N$ and $N_b$ respectively.

The choice of the above specific forms for the Liouvillians is physically motivated, especially by applications in microwave and magnonic technologies~\cite{Burin2015,Klimov2018,Goetz2017,Schloer2019,Burnett2019,Burnett2014,deGraaf2018,Mueller2019,Tabuchi2014,Pfirrmann2019,LachanceQuirion2019,Woltersdorf2009,Mihalceanu2018,MaierFlaig2017}. When applied to these systems, the bosonic modes $\hat{s}_n$ represent electromagnetic degrees of freedom, and our model recovers the usual cavity quantum electrodynamics description under the rotating wave and independent TLS approximation~\cite{CohenTannoudji1992Book,MeystreBook}. In other words, our model is appropriate to describe these systems provided that the validity conditions for the rotating wave approximation are fulfilled both for the coherent driving and for the system-TLS interaction terms, namely $\vert\omega_n-\omega_d\vert \ll \omega_n+\omega_d$, $\vert\omega_{Bi}-\omega_d\vert \ll \omega_{Bi}+\omega_d$, $\vert\omega_n-\omega_{Bi}\vert \ll \omega_n+\omega_{Bi}$, $\vert\coupl{in}\vert \ll \omega_n+\omega_{Bi}$, $\vert\On\vert \ll \omega_n+\omega_d$ and $\vert\Obi\vert \ll \omega_{Bi}+\omega_d$. Although these conditions are usually fulfilled, the rotating wave approximation could break down e.g. in strongly coupled systems~\cite{CohenTannoudji1992Book,MeystreBook}. Similarly, the independent TLS approximation, namely neglecting any direct (e.g. dipole-dipole) coupling between the TLS, is also well justified since the TLS densities in these systems are typically sufficiently small. Finally, in these platforms the drivings of the system and the TLS usually have the same frequency $\omega_d$, as they originate from the same microwave signal.
For all the above reasons, our model can be directly  applied to most microwave and magnonic platforms.

\subsection{Born-Markov master equation}
\label{sec:derivation}

Our aim is to trace out the TLS degrees of freedom and obtain an effective equation of motion for the bosonic system. First, we transform the Liouville--von Neumann equation \eqref{Liouville-von-Neumann-equation} to a frame rotating at the driving frequency $\wdr$, by applying the unitary transformation
\be
\begin{split}\label{U-rotating-frame}
	\Urf&=\exp\spare{ \im\wdr t \pare{\sum_n \sopd_n \sop_n + 
	\sum_{i}\frac{\sgm{zi}}{2} } }.
\end{split}
\ee
The Liouville--von Neumann equation in the rotating frame is 
\be\label{LvN-eq-rf}
    \dotrhotot\rf = \pare{\curlyL\rf_S + \curlyL\rf_B + \curlyL\rf_I } [\rhotot\rf], 
\ee
with $\rhotot\rf(t) = \Urf \rhotot (t)\Urfd$, and time-independent Liouvillian superoperators for the system, the bath and the interaction part, given respectively as $\curlyL\rf_S,\curlyL\rf_B,\curlyL\rf_I$. These three Liouvillians have the same form as Eqs.~\eqref{system-liouvillian}, \eqref{bath-Liouvillian}, and \eqref{interaction-Liouvillian} with modified, time-independent system and bath Hamiltonians given by
\be\label{HamiltonianS-rf}
    \Hop\rf_S = \hbar\sum_n \cparelb \dn\sopd_n\sop_n + \pare{\On \sop_n + \hc }\cparerb,
\ee
\be\label{HamiltonianB-rf}
    \Hop\rf_B = \frac{\hbar}{2} \sum_{i} \cparelb\dbi\sgm{zi} + \pare{\Obi \sgm{+i}  + \hc }\cparerb,
\ee
with $\dn=\wn-\wdr$ and $\dbi=\wbi-\wdr$. 

Second, we make the following change of variables in the interaction Liouvillian $\curlyL\rf_I$: 
\be\label{sigmatilde-def}
    \sgmt{\alpha i}(t) = \sgm{\alpha i} - \mv{\sgm{\alpha i}}(t) \hspace{0.2cm};\hspace{0.2cm}\left(\alpha=+,-,z\right),
\ee
where $ \mv{\sgm{\alpha i}}(t)=\tr [ \sgm{\alpha i} \rhotot\rf(t)]$. From \eqnref{interaction-Liouvillian}, we obtain an effective system driving term such that \eqnref{HamiltonianS-rf} acquires a total, time-dependent driving given by the rate
\be\label{effective-driving}
    \On'(t) = \On +\sum_{i} \coupl{in} \mv{\sgm{+i}}(t).
\ee
This time-dependent term is typical for driven baths, and it can be interpreted as an effective force acting on the system~\cite{Reichert2016,Grabert2018,Grabert2016}.

Finally, we transform \eqnref{LvN-eq-rf} to a second rotating frame, by applying the unitary transformation 
\be\label{U-int-pic}
	\Uint = \exp \spare{\im t \pare{\sum_n \dn\sopd_n\sop_n + \frac{\Hop\rf_B}{\hbar} } }. 
\ee
The Liouville--von Neumann equation in the second rotating frame is
\be
\begin{split}\label{Liouville-von-Neumann-equation-interaction-picture}
	\dotrhotot\intpic &= \cpare{\curlyL_S\intpic(t) + \curlyB\intpic(t) +\curlyL_I\intpic(t)}[\rhotot\intpic]\\
	&\equiv \curlyL\intpic(t) [\rhotot\intpic],
\end{split}
\ee
with $\rhotot\intpic(t) = \Uint \rhotot\rf(t) \Uintd$, and $\curlyL\intpic(t)$ denoting the total Liouvillian. Here the system Liouvillian is given by
\be \label{Liouvillian-system-int}
\begin{split}
	&\curlyL_S\intpic(t) [\rhoop] = -\frac{\im}{\hbar}\sum_n [\On'(t)\hat{s}_n\intpic(t) + \hc, \rhoop] +\curlyS[\rhoop],
\end{split}
\ee
with $\hat{O}\intpic(t) = \Uint \hat{O} \Uintd$. The bath Liouvillian $\curlyB\intpic(t)$ maintains the same form as \eqnref{bath-dissipation}, under the substitution $\sgm{\alpha i}\to\sgm{\alpha i}\intpic(t)$. Finally, the interaction Liouvillian reads
\be \label{Liouvillian-interaction-int}
\begin{split}
    \curlyL_I\intpic(t) [\rhoop] &= -\frac{\im}{\hbar}[\Hint\intpic(t), \rhoop] \\
    &= -\frac{\im}{\hbar}\sum_{n i} \spare{\coupl{in}\sgmt{+i}\intpic(t)\sop_n\intpic(t) + \hc, \rhoop}.
\end{split}
\ee

The next step in the master equation derivation is to trace out the TLS bath in the Born-Markov approximation. Since the standard approach assumes that the system and the bath are closed and this is not our case, we employ a generalized approach based on projection superoperator techniques~\cite{BreuerPetruccione}. We define a projection superoperator $\mathcal{P}$ that acts on the space of density matrices for the compound system and bath in the following way,
\be\label{Projector-definition}
    \curlyP \rhoop(t) = \tr_B [\rhoop (t)] \otimes \rhoBs\intpic(t),
\ee
where $\tr_B$ indicates a partial trace over the bath degrees of freedom, and $\rhoBs\intpic(t)$ denotes the stationary state of the bath Liouvillian, namely
\be\label{BathStationaryState}
    \curlyB\intpic(t)[\rhoBs\intpic(t)]=\curlyL_B\rf[\rhoBs\rf]=0,
\ee
with $\rhoBs\rf = \Uintd \rhoBs\intpic(t) \Uint$. Note that this stationary state of the bath is not thermal due to the presence of driving. The evolution of the projection $\curlyP \rhotot\intpic(t)$ can be expressed by the Nakajima-Zwanzig equation~\cite{BreuerPetruccione}:
\begin{multline}\label{Nakajima-Zwanzig-equation}
	\fd{}{t} \curlyP \rhotot\intpic(t) = \curlyP \curlyL\intpic(t) \curlyP \rhotot\intpic(t)\\
	+\curlyP \curlyL\intpic(t) \integral{0}{t}{\tvar}
	\curlyG(t,\tvar)
	\curlyQ 
	\curlyL\intpic(\tvar) \curlyP \rhotot\intpic(\tvar), 
\end{multline}
where $\curlyQ = 1-\curlyP$,  $\curlyG(t,\tvar) = \mathcal{T}_+\exp\spares{\integral{\tvar}{t}{\tvarp}\curlyQ\curlyL\intpic (\tvarp)}$, with $\mathcal{T}_+$ the time-ordering superoperator, and we assume system and bath are uncorrelated at $t=0$, i.e., $\curlyQ \rhotot\intpic(0) = 0$. Equation~\eqref{Nakajima-Zwanzig-equation} is an exact reformulation of \eqnref{Liouville-von-Neumann-equation-interaction-picture}, and it is particularly convenient for two reasons. First, its form allows us to perform the Born-Markov approximation in a simpler fashion (see below). Second, by taking the trace of \eqnref{Nakajima-Zwanzig-equation} over the bath degrees of freedom, one directly obtains a dynamical equation for the reduced density matrix of the system.

We now simplify \eqnref{Nakajima-Zwanzig-equation} with the Born and Markov approximations. First, we assume the TLS bath is not significantly affected by its interaction with the system, a condition which is satisfied when the system-bath interaction is weak, namely $\vert\coupl{in}\vert\ll\kti$. This allows us to undertake the Born approximation, 
\be\label{BornApprox}
    \rhotot\intpic(t) \approx \rhoBs\intpic(t)\otimes\rhoS\intpic(t),
\ee
where the bath stationary state is given by \eqnref{BathStationaryState}, and $\rhoS\intpic(t)=\tr_B [\rhoop\intpic (t)]$ is the reduced density matrix of the system. Under the Born approximation the expectation values of the Pauli operators, defined below \eqnref{sigmatilde-def}, and hence the effective driving rate \eqnref{effective-driving}, become time-independent. Moreover, under this approximation $\curlyP \curlyL_I\intpic(t)  \curlyP = 0$, allowing us to largely simplify the Nakajima-Zwanzig equation~\cite{BreuerPetruccione}. Second, we assume that the two-time correlation functions of the bath operators appearing in the interaction Liouvillian, \eqnref{Liouvillian-interaction-int}, decay on a much faster time scale than any characteristic time scale of the system Liouvillian $\curlyL_S\intpic(t)$~\footnote{It is at this point where the change of variables in \eqnref{sigmatilde-def} becomes crucial. Indeed, if the bath operators appearing in the interaction Liouvillian had non-zero expectation value, some of their two-time correlation functions would not decay, preventing us from performing the Markov approximation. }. This assumption is valid provided that $\kti\gg\gamma_n,\vert\On'\vert$, and it allows us to undertake the Markov approximation, namely to substitute
\be\label{MarkovApprox}
    \rhotot\intpic(\tvar) =  \rhoBs\intpic(\tvar)\otimes\rhoS\intpic(\tvar)\approx \rhoBs\intpic(\tvar)\otimes\rhoS\intpic(t)
\ee
in the integrand of \eqnref{Nakajima-Zwanzig-equation} and take the upper integration limit to infinity~\cite{BreuerPetruccione}.

Within the Born-Markov approximation we obtain the following master equation for the system dynamics, 
\begin{multline}\label{Born-Markov-master-equation}
	\fd{}{t} \rhoS\intpic(t) = \curlyL_S\intpic(t) [\rhoS\intpic (t)] - \integral{0}{\infty}{\tvar} \tr_B \\
	\left[\Hint\intpic(t),\tilde\curlyG(t,\tvar)\left[\Hint\intpic(t-\tvar),\rhoBs\intpic(t-\tvar) \otimes \rhoS\intpic(t)\right]\right],
\end{multline}
where $\tilde\curlyG(t,\tvar) = \mathcal{T}_+\exp\spares{\integral{t-\tvar}{t}{\tvarp}\curlyB\intpic (\tvarp)}$.  \eqnref{Born-Markov-master-equation} is a generalisation of the standard Born-Markov master equation. Note that for a closed ($\curlyB =0$)  and undriven ($\rhoBs\intpic(t)$ time-independent) TLS bath, \eqnref{Born-Markov-master-equation} reduces to the usual form more commonly used in simpler open quantum systems~\cite{BreuerPetruccione}. The second term in \eqnref{Born-Markov-master-equation} captures the TLS-induced dissipation, and it is completely determined by two-time correlation functions of Pauli operators~\cite{Carmichael}, 
\begin{multline}\label{bath-traces}
    \tr_B \spare{\sgmt{\pm i}\intpic(t) \tilde\curlyG(t,\tvar)  \cpare{\sgmt{\pm j}\intpic(t-\tvar)\rhoBs\intpic(t-\tvar)} }\propto\delta_{ij},
\end{multline}
where $\delta_{ij}$ follows from the independent TLS assumption. 

Expanding \eqnref{Born-Markov-master-equation}, the final Born-Markov master equation can be written, in the frame rotating at frequency $\omega_d$, as
\be \label{master-equation}
\begin{split}
    \dotrhoS\rf = -\im [\Hop_S', \rhoS\rf] +\curlyS[\rhoS\rf]+\curlyS_\text{TLS}[\rhoS\rf].
\end{split}
\ee
The term $\Hop_S'$ is given by
\begin{multline}\label{master-equation-TLS-contributions-H}
    \Hop_S' = 
	 \hbar\sum_n    \cparelb \dn        \sopd_n\sop_n + \pare{\On' \sop_n  + \hc }\cparerb \\
	+\hbar\sum_{mn} \cparelb \delta_{mn}\sopd_m\sop_n + \pare{\sqrate_{mn} \sop_m\sop_n  + \hc }\cparerb, 
\end{multline}
and it includes the original system Hamiltonian plus additional coherent dynamics induced by the TLS bath. These dynamics include, first, a modified system driving rate given by \eqnref{effective-driving} in the Born approximation, namely
\be \label{master-equation-effective-driving}
\begin{split}
	\On'&= \On +\sum_i \coupl{in} \mv{\sgm{+i}}\sts,
\end{split}
\ee
where $\mv{\sgm{\alpha i}}\sts=\tr_B [ \sgm{\alpha i} \rhoBs\rf]$. Second, a frequency shift of each system mode and interactions of the beam-splitter type between different system modes ($\delta_{mn}$), and third, interactions of the two mode-squeezing type ($\sqrate_{mn}$), given by the rates
\be\label{deltamn-def}
    \delta_{mn}   = -\frac{\im}{2} \pare{ \GammaB^{mn} + \GammaC^{mn} } +\frac{\im}{2} \pare{ \GammaB^{nm} + \GammaC^{nm} }^*,
\ee
\be\label{gmn-def}
    \sqrate_{mn}  = -\frac{\im}{2} \spare{\GammaA^{mn} - (\GammaD^{nm})^* }.
\ee
The above rates are expressed in terms of the TLS one-sided power spectral densities
\begin{multline} \label{master-equation-Gammas}
    \Gamma_{\alpha\beta}^{mn} = \sum_i \coupl{in}^{(\alpha)}\coupl{im}^{(\beta)}     
    \\\times
    \integral{0}{\infty}{\tvar} \mv{ \sgmt{\alpha i}(\tvar)\sgmt{\beta i}(0) }\sts
    e^{\beta\im\dm \tvar},
\end{multline}
where $\alpha,\beta = \pm$, $\coupl{in}^{(+)} = \coupl{in}$ and $\coupl{in}^{(-)}=\coupl{in}^*$, and
\begin{multline} \label{TTCF}
    \mv{ \sgmt{\alpha i}(\tvar)\sgmt{\beta i}(0) }\sts=\tr_B \spare{\sgmt{\pm i} e^{\curlyL_B\rf \tvar} \cpare{\sgmt{\pm j}\rhoBs\rf} } 
\end{multline}
are the two-time correlation functions in the bath steady state. As $\rhoBs\rf$ is time-independent, the correlators \eqnref{TTCF} do not depend explicitly on time $t$. The last term in \eqnref{master-equation} describes the dissipative dynamics induced by the TLS bath, and it takes the most general form possible for quadratic Lindblad dissipators,
\begin{multline}\label{master-equation-TLS-contributions-S}
    \curlyS_\text{TLS}[\rhoS] =  
	\sum_{mn}  \cparelb 
	\pare{\diffrate_{mn}  \curlyD_{\sop_m,\sop_n} + \hc }\\
	+
	\gammaB^{mn} \curlyD_{\sopd_m,\sop_n}[\rhoS] + \gammaC^{mn} \curlyD_{\sop_m,\sopd_n}[\rhoS]\cparerb.
\end{multline}
The rates in the above equation are given by
\be\label{gammamn-definition}
    \gammaBC^{mn} = \GammaBC^{mn} +(\GammaBC^{nm})^* ,
\ee
and
\be\label{Gammamn-definition}
    \diffrate_{mn} = \GammaA^{mn} + (\GammaD^{nm})^*.
\ee

The power spectral densities Eq.~\eqref{master-equation-Gammas}, and hence all the rates in the master equation, can be calculated analytically as a function of the parameters $N, \Obi, \wbi,\wdr,\kdci,\kdpi,\wn$ and $T$. Using the optical Bloch equations and the quantum regression theorem~\cite{Carmichael}, we compute the integral
\begin{multline}\label{TTCF-integrals} 
	\integral{0}{\infty}{\tvar} \mv{ \sgmtv(\tvar)\sgmt{\pm i}(0) }\sts 
	\hspace{1mm} e^{\pm\im\dm \tvar}=\\
	=-(A_i\pm\im\dm \mathds{1}_3)^{-1} \mv{ \sgmtv\sgmt{\pm i} }\sts, 
\end{multline}
where $\sgmtv = (\sgmt{+i},\sgmt{-i},\sgmt{zi})^T$, $\mv{ \sgmtv\sgmt{\pm i} }\sts = \tr [ \sgmtv\sgmt{\pm i}\rhoBs\rf ]$, $\mathds{1}_d$ is the $d-$dimensional identity matrix, and $A_i$ is a coefficient matrix given by
\be\label{OBE-Amatrix}
    A_i=
    \begin{pmatrix}
 	\im\dbi-\kti& 0&              -\im\Obi^*/2 \\
 	0&          -\im\dbi-\kti&    \im\Obi/2 \\
 	-\im\Obi&    \im\Obi^*&       -\kdci[1+2\bar n(\wbi)]
    \end{pmatrix}.
\ee 
Introducing \eqnref{TTCF-integrals} in \eqnref{master-equation-Gammas} and using the commutation relations of the Pauli operators, we obtain analytical expressions for the rates $\Gamma_{\alpha\beta}^{mn}$ in terms of the spectral properties of the matrix $A_i$ and of the stationary-state expectation values of the Pauli operators, 
\begin{align}
    \mv{\sgm{+,i}}\sts &= \frac{-1}{1+2\bar n(\wbi) + s_i}
    \frac{\Obi^*}{2(\dbi+\im \kti)},\label{sigmaP-meanvalue}\\
    \mv{\sgm{z,i}}\sts &= \frac{-1}{1+2\bar n(\wbi) + s_i},\label{sigmaZ-meanvalue}
\end{align}
given in terms of the saturation parameter
\be \label{saturation-parameter}
s_i =  \frac{\kti}{\kdci}\frac{\vert\Obi\vert^2 }{ \kti^2+\dbi^2 }.
\ee
The master equation \eqnref{master-equation} is the main result of this work.

We conclude this section with several remarks. First, note that due to the Born-Markov approximation the master equation models an environment of lossy TLS ($\kti \gg \vert\coupl{in}\vert,\gamma_n,\vert\On'\vert$), and it is thus adequate to model baths of near-resonant solid state impurities which typically have large linewidths~\cite{Burin2015,MaierFlaig2017,Woltersdorf2009,Andersson2021,Pfirrmann2019,TenorioPearl2017}. However, although our initial model is formally identical to usual cavity quantum electrodynamics, the derived master equation is not appropriate to describe typical quantum optics cavity QED setups where the TLS (atoms, quantum dots, etc.) are characterized by a very narrow linewidth $\kti$. Second, note that within Born-Markov approximation it is consistent to consider the TLS to be in their stationary state, defined by \eqnref{BathStationaryState}, already at the time $t=0$, i.e. right after the driving is turned on. Indeed, since the time scale at which the TLS evolve to their stationary state (given by $\kti^{-1}$) is much shorter than the time scale of the system dynamics (given by $\gamma_n{}^{-1}, \vert\On'\vert^{-1}$), this relaxation can be considered instantaneous. Third, note that the most unconventional effective dynamics in the master equation originate exclusively from the driving of the TLS bath. In the absence of this driving, i.e., at $\Omega_{Bi}=0$, one has $\On' = \On$ and $\diffrate_{mn}=\sqrate_{mn}=0$, and the master equation recovers the usual form obtained for simpler baths (e.g. undriven bosonic baths), containing only absorption, decay, and particle-conserving interactions. Finally, note that by following the above steps our derivation can be directly extended to include the following additional features: (i) rotating terms ($\propto\sgm{-i}\sop_n$ and/or $\propto\sgm{zi}\sop_n$) in the interaction Liouvillian \eqnref{interaction-Liouvillian}. (ii) Different driving frequencies $\omega_{d,S}$ and $\omega_{d,B}$ for the system and the bath. (iii) Arbitrarily strong system driving $\Omega_n$~\footnote{This can be done by transforming to a different frame, this time with respect to a Hamiltonian including the system driving. A more complicated, but still analytical master equation can be derived in this case.} and/or an arbitrary system dissipator $\curlyS$, as long as their associated time scales are consistent with the Markov approximation. (iv) A bath composed of multilevel systems with dissipators $\curlyB$ of an arbitrary form, as long as the time scales of these dissipators are consistent with the Born and Markov approximations.

\section{TLS-induced dynamics: amplification, squeezing and system instabilities}
\label{sec:case-study}

In this section we focus on understanding the effective system dynamics induced by the driven TLS bath. In \secref{sec:particular-case} we introduce the particular case of a single bosonic mode coupled to a bath of $N$ identical TLS. In \secref{sec:rates} we characterize the rates appearing in the master equation for this particular case. In \secref{sec:dynamics} we characterize the steady-state of the system and the exotic properties arising from the non-thermal bath. 

\subsection{Particular case: single mode coupled to a bath of identical TLS}
\label{sec:particular-case}

\begin{table}[t]
	\caption{Summary of the parameters in the particular case of a single bosonic mode coupled to a bath of $N$ identical TLS.}
	\label{tab:parameters}
	\begin{ruledtabular}
		\begin{tabular}{ll}
			Parameter & Definition\\
			\hline
			$\wdr$          & driving frequency\\
			$\wsingle$	    & system frequency\\
			$\dsingle$ 		& system detuning,\\
			$\Osingle$ 		& external driving rate of the system\\
			$\gamma_0$ 		& intrinsic decay rate of the system\\
			\hline
			$\wb$ 			& TLS frequency\\
			$\db$ 	    	& TLS detuning\\
			$\Ob$ 			& external driving rate of the TLS\\
			$\kdc$	        & TLS decay rate\\
			$\kdp$ 			& TLS dephasing rate\\
			$\kt$           & TLS transverse decay rate\\
			$s$             & TLS saturation parameter\\
			\hline
			$\coup$ 		& system-TLS coupling rate\\
			$\Oeff$ 		& TLS-induced effective driving rate of the system\\
			$\gammaBC$      & TLS-induced system absorption/emission rate\\
			$\gamma$        & TLS-induced decay rate\\
			$\sqrate,\diffrate$ & TLS-induced system squeezing rates\\
			$\delta$        & TLS-induced system frequency shift
		\end{tabular}
	\end{ruledtabular}
\end{table}

To illustrate the rich phenomenology induced by the driven TLS bath, hereafter we focus on the particular case of a single bosonic mode, coupled to a bath of $N$ identical TLS. The bosonic mode is described by creation and annihilation operators $\sopd$ and $\sop$. Its free dynamics, given by $\curlyL_S$ in \eqnref{system-liouvillian}, is characterized by a frequency $\wsingle$, a driving rate $\Osingle$, and an intrinsic decay rate $\gamma_0$. 
Since the TLS are identical, all the rates become TLS-independent, i.e., $\{\wbi,\kdci,\kdpi,\Obi,\coupl{in},\dbi,\kti,s_i\} \rightarrow \{\wb,\kdc,\kdp,\Ob,\coup,\db,\kt,s\}$. 
Similarly, since this particular case consists of a single mode, for simplicity we denote its TLS-induced master equation rates $\{\delta_{nn},\sqrate_{nn},\gammaBC^{nn},\diffrate_{nn}\} \rightarrow \{\delta,\sqrate,\gammaBC,\diffrate\}$. 
Without loss of generality we assume $\Ob \in \mathbb{R}$. All the parameters appearing in this particular case are summarised in \tabref{tab:parameters}. The master equation in this case has the same form as \eqnref{master-equation}, with a simplified Hamiltonian
\be \label{one-mode-terms1}
    \Hop_S' = \dLamb\sopd\sop + (\Oeff\sop +\hc) + (\sqrate\sop^2 + \hc),
\ee
and a dissipator
\begin{multline} \label{one-mode-terms2}
    \curlyS_\text{TLS}[\rhoS] =  \gammaB \curlyD_{\sopd, \sop} [\rhoS] 
    + \gammaC \curlyD_{\sop, \sopd} [\rhoS] \\
    + \pare{ \diffrate \curlyD_{\sop,\sop} [\rhoS] + \hc }.
\end{multline}
Here $\dLamb = \dsingle+\delta$, with the system detuning $\dsingle = \wsingle-\omega_d$, and $\Oeff = \Osingle + N \coup \mv{  \sgm{+ i}}$. 

From the master equation \eqnref{master-equation} one can obtain a dynamical equation for the expectation value of any system operator $\hat{S}$, namely ${\mathrm{d} \mv{\hat {S}} }/{\mathrm{d}t} = \tr [\hat{S}\dotrhoS\rf]$. Since the master equation is quadratic, the expectation values of the first- and second-order momenta $\bm{v}=(\mv{\sopd \sop},\mv{\sop}, \mv{\sopd},\mv{\sop^2},  \mv{(\sopd)^2})^T$ obey a closed linear system of differential equations of the form $\dot{\bm{v}} =A_S  \bm{v}+\bm{a}_S$, with
\be \label{sds-system}
A_S =\begin{pmatrix}
    -(\gamma_0 +\gamma)&\im\Oeff&-\im\Oeff^*&2\im\sqrate&-2\im\sqrate^*\\
    0&\tilde\Delta^*&-2\im\sqrate^*&0& 0\\
    0&2\im\sqrate&\tilde\Delta&0&0\\
    -4\im\sqrate^*&-2\im\Oeff^*&0&2\tilde\Delta^*&0\\
     4\im\sqrate&0&2\im\Oeff&0&2\tilde\Delta\\
\end{pmatrix}
\ee
and
\be\label{sds-independentvector}
    \bm{a}_S =\left(\gammaB,-\im \Oeff^*, \im \Oeff,-2\im \sqrate^* -\diffrate^*,2\im \sqrate -\diffrate \right)^T.
\ee
In the above equations we define $\tilde\Delta = \im\dLamb-(\gamma_0 +\gamma)/2$, where the important parameter $\gamma = \gammaC-\gammaB$ is the TLS-induced decay rate of the system. In the following we analyze the rates and the effective dynamics induced by the TLS in this particular scenario.

\subsection{Master equation rates: Mollow sidebands and amplification}
\label{sec:rates}

\begin{table}[t]
	\caption{Default values chosen for the parameters in all the figures throughout the text. All the rates are expressed in terms of the TLS frequency $\wb$.}
	\label{tab:numbers}
	\begin{ruledtabular}
		\begin{tabular}{ll}
			Parameter & Value\\
			\hline
			temperature 				& $T=0$\\
			system-TLS coupling rate	& $\coup = 10^{-8} \wb$\\
			number of the TLS 			& $N=10^5$\\
			TLS decay rate 				& $\kdc = 10^{-4} \wb$\\
			TLS dephasing rate 			& $\kdp =0$\\
			system decay rate			& $\gamma_0 = 10^{-7} \wb$\\
		\end{tabular}
	\end{ruledtabular}
\end{table}

We divide this section into three blocks corresponding to the effective system driving rate $\Oeff$, the rates associated with particle-non-conserving terms $\diffrate$ and $\sqrate$, and the decay rate and frequency shift $\gamma$ and $\delta$.

\subsubsection{Effective driving rate \texorpdfstring{$\Oeff$}{}}
\label{sec:driving}

We first consider the effective driving rate $\Oeff$. For simplicity we assume the system is not independently driven, i.e., $\Osingle = 0$, so that $\Oeff = N \coup \mv{ \sgm{+ i}}$. This rate originates from the expectation value of the TLS operators. Its effect on the system is analogous to the \textit{coherent} electromagnetic scattering of a driven TLS, namely to the coherent part of the resonance fluorescence spectrum~\cite{CohenTannoudji1992Book,MeystreBook}. In the context of resonance fluorescence, the coherent scattering results from the emission of light by the classical component of the TLS dipole moment. Indeed, the square modulus of the effective driving rate reads 
\be\label{Omega0prime}
    \vert\Oeff\vert^2 = (N\vert \coup \vert)^2\frac{\kdc}{4\kt}\frac{  s }{   (1+s)^2 }, 
\ee
where $s$ is the saturation parameter \eqnref{saturation-parameter} characterizing the regimes of strong $(s\gg1)$ and weak $(s\ll 1$) TLS driving. Equation~\eqref{Omega0prime} has an identical dependence on the saturation parameter as the coherently scattered power in resonance fluorescence~\cite{CohenTannoudji1992Book,MeystreBook}. 

\begin{figure}[t]
	\centering
	\includegraphics[width=\linewidth]{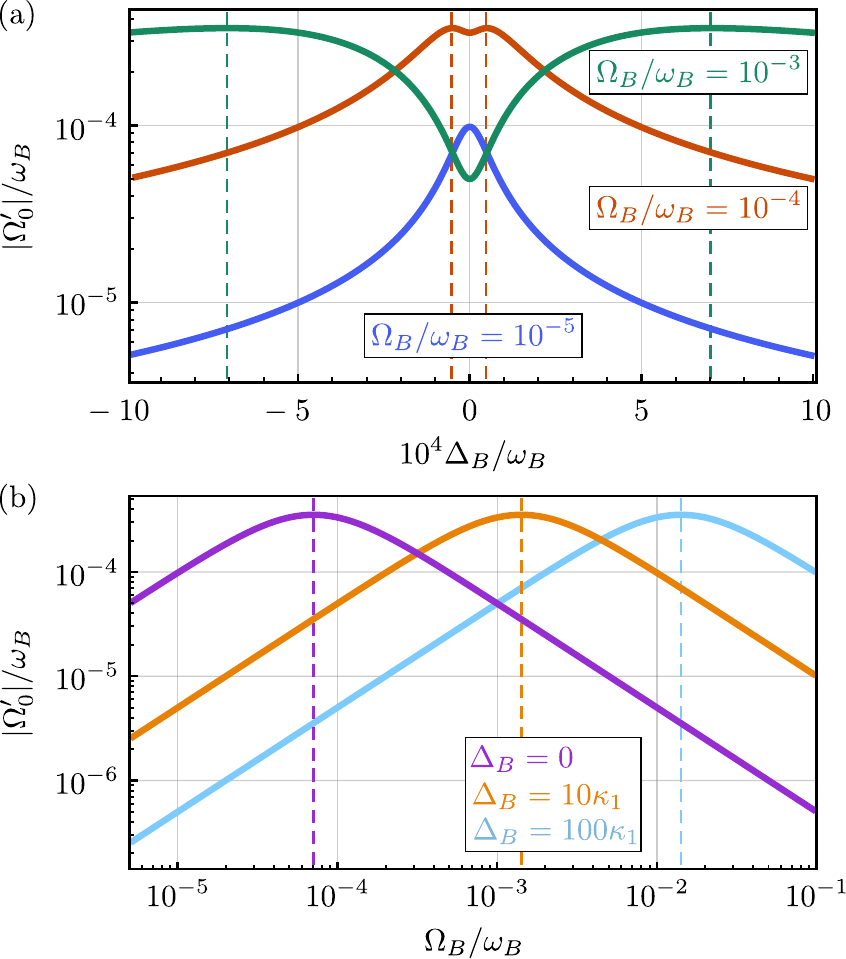}
	\caption{
	(a) Dependence of the effective driving rate $\Oeff$ on the detuning between the TLS and their driving, $\db$, for different values of the TLS driving rate $\Ob$. In the red and the green curves, vertical dashed lines indicate the optimum detuning $\Delta_{B,\rm opt}=\pm\sqrt{\vert\Ob\vert^2/2-\kappa_t^2}$ (see main text).
	(b) Dependence of the effective driving rate $\Oeff$ on TLS driving rate $\Ob$ at fixed detunings $\db$. The dashed lines mark the value of the driving rate at which $s=1$. 
	}
	\label{fig:driving}
\end{figure}

The effective driving rate $\Oeff$ is shown in \figref{fig:driving}(a) as a function of detuning between the TLS and their driving, $\db$, and for different values of the TLS driving rate $\Ob$. The remaining parameters, chosen as in Table~\ref{tab:numbers}, are consistent with our approximations and, to our knowledge, with typical values in microwave and magnonics platforms. At weak driving $\Ob \ll \kt$ (blue curve) the saturation parameter is always low, $s\ll 1$ (see Eq.~\eqref{saturation-parameter}) and the effective driving rate has a single peak at $\db=0$, as $\vert\Oeff\vert \sim \sqrt{s}$. As the driving rate reaches values comparable to the TLS linewidth, $\Ob\sim\kt$, the saturation parameter can become $s \gtrsim 1$. In this regime the effective driving rate acquires a double-peaked structure as shown by the red and green curves in \figref{fig:driving}(a). This is a characteristic indication of the energy levels of the TLS becoming dressed by the strong driving, with eigenenergies $\pm\hbar \sqrt{\db^2+\vert\Ob\vert^2}/2$~\footnote{This can be readily checked by diagonalizing the TLS Hamiltonian, \eqnref{bath-Hamiltonian}, in the frame rotating at the driving frequency.}. As a consequence of this shift in the TLS bare energies, the driving is not resonant anymore at $\db=0$, and the conditions for maximum scattering shift to detunings $\Delta_{B,\rm opt}^2 = \vert\Ob\vert^2/2-\kappa_t^2\ne 0$, marked by vertical dashed lines in \figref{fig:driving}(a). The above argument is confirmed by \figref{fig:driving}(b), where we show the effective driving rate at fixed detuning $\db$ as a function of driving rate $\Ob$. At weak driving it grows linearly whereas at strong driving it linearly decays, $\vert \Oeff\vert \propto 1/\sqrt{s} \propto 1/\Ob$, as the fixed driving frequency becomes increasingly off-resonant with respect to the energy gap between the dressed TLS states. The maximum of $\Oeff$ is attained at $s=1$ (see \eqnref{Omega0prime}), marked by vertical dashed lines in \figref{fig:driving}(b). As the detuning $\db$ increases, the condition $s=1$ is attained at larger driving rates $\Ob$, and thus the curves in \figref{fig:driving}(b) shift forward horizontally.

\subsubsection{Squeezing rates \texorpdfstring{$\diffrate$ and $\sqrate$}{} }
\label{sec:squeezing}

All the remaining rates in Eqs.~\eqref{one-mode-terms1}-\eqref{one-mode-terms2}, namely $\gamma,\delta,\diffrate,$ and $\sqrate$, originate from the fluctuations of the TLS bath, i.e. from the Fourier transform of the two-time correlation functions $\mv{ \sgmt{\alpha i}(t+\tau)\sgmt{\beta i}(t) }$ with $\alpha,\beta = \pm$, evaluated at the system frequency $\dsingle$ (see \eqnref{master-equation-Gammas}). These rates and their associated effective dynamics thus show some properties similar to the \textit{incoherent} scattering spectrum of resonance fluorescence as we will see below. Here we focus on the two rates $\diffrate$ and $\sqrate$ which arise from correlators $\mv{ \sgmt{\alpha i}(t+\tau)\sgmt{\alpha i}(t) }$. They represent a contribution to the system effective dynamics exclusively induced by the driving of the bath, as they vanish at $\Ob=0$. 
Both $\diffrate$ and $\sqrate$ appear in excitation-non-conserving terms in the master equation and can induce squeezing and instabilities on the system as we will see below. As evidenced by Eqs.~\eqref{sds-system}-\eqref{sds-independentvector}, the dissipator associated to $\diffrate$ only affects the steady state of the system while leaving any dynamical rate unchanged. 

\begin{figure}
	\centering
	\includegraphics[width=\linewidth]{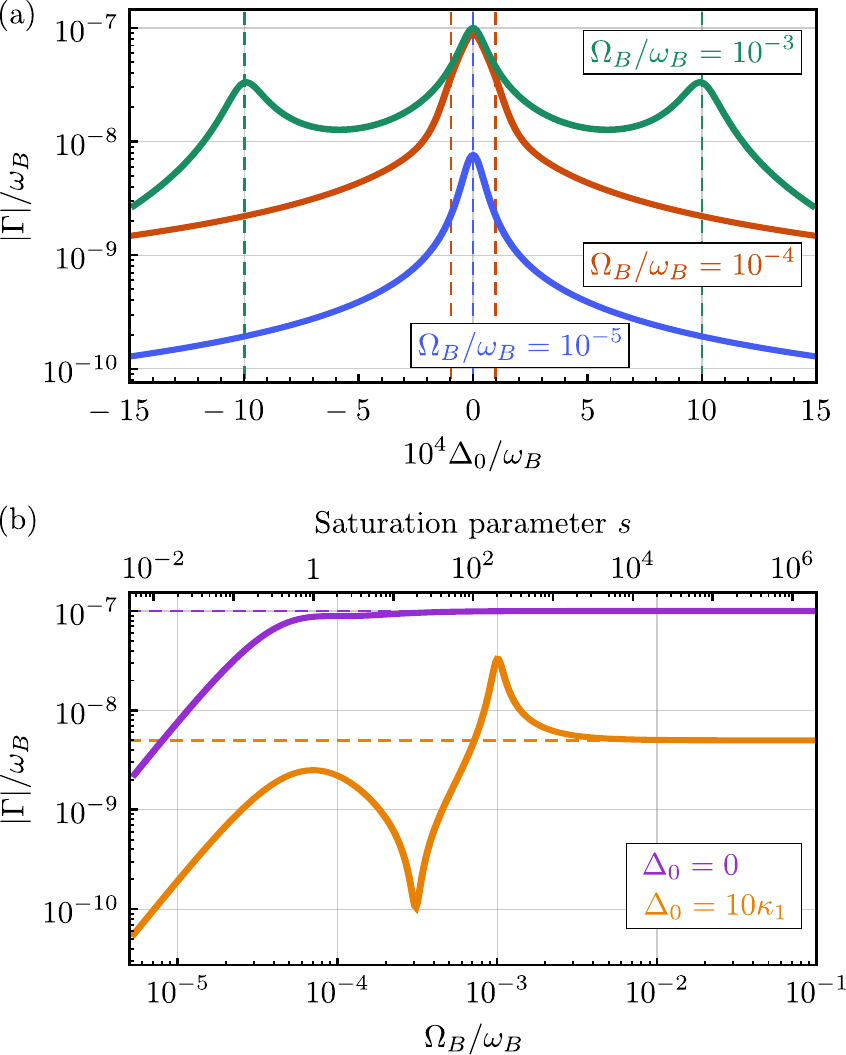}
	\caption{
		(a) Dependence of the rate $\diffrate$ on the system detuning $\dsingle$ for $\db=0$, different TLS driving rates $\Ob$, and the parameters in Table~\ref{tab:numbers}. The dashed lines denote the frequencies of the Mollow sidebands (see main text). 
		(b) Dependence of the rate $\diffrate$ on TLS driving rate $\Ob$ for different system detunings $\dsingle$ and the same parameters as in panel (a). The horizontal dashed lines denote the strong-driving limit \eqnref{limit-diffrate}.
		}
	\label{fig:dissip-sq-rate}
\end{figure}

\begin{figure}
	\centering
		\includegraphics[width=\linewidth]{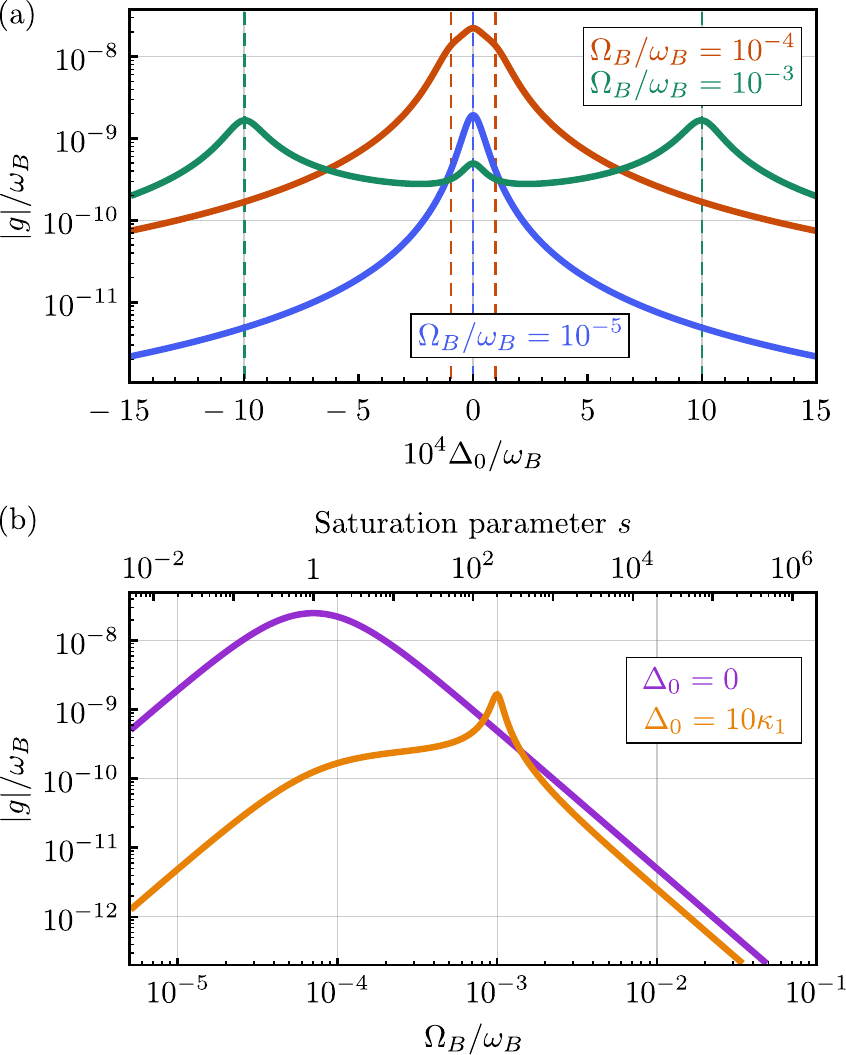}
		\caption{
		(a) Dependence of the rate $\sqrate$ on the system detuning $\dsingle$ for $\db=0$, different TLS driving rates $\Ob$, and the parameters in Table~\ref{tab:numbers}. The dashed lines denote the frequencies of the Mollow sidebands (see main text). 
		(b) Dependence of the rate $\sqrate$ on TLS driving rate $\Ob$ for different system detunings $\dsingle$ and the same parameters as in panel (a).
		}
		\label{fig:squeezing-rate}
\end{figure}

In the following we focus on the case of resonantly driven TLS, namely $\wb=\wdr$ ($\db=0$).
The rates $\diffrate$ and $\sqrate$ are displayed in \figref{fig:dissip-sq-rate}(a) and \figref{fig:squeezing-rate}(a) respectively, as a function of the system detuning $\dsingle$ for three values of the driving rate $\Ob$ and the parameters in Table~\ref{tab:numbers}. In this regime, the following analytical expressions can be derived,
\begin{multline}\label{squeezing-analytical}
    \left(\begin{array}{c}
         \sqrate  \\
         \diffrate 
    \end{array}\right)= \frac{N G^2}{2\kdc} \frac{-s}{(1+s)^2f(s,\dsingle/\kdc)}
    \\
    \times\left(\begin{array}{c}
         \im(\im\dsingle/\kdc-1)(1+s)  \\
         s^2+2s+4(\im\dsingle/\kdc-1)^2 
    \end{array}\right),
\end{multline}
with $f(s,d) = \spares{s+2(\im d-1)(\im d-1/2)}\pares{\im d-1/2}$. At low saturation $s\ll1$ (blue curves in \figref{fig:dissip-sq-rate}(a) and \figref{fig:squeezing-rate}(a)) the rates have a Lorentzian profile, as $\diffrate,g\sim s$. At high saturation $s\gg 1$ (green curves) two additional side peaks emerge. This is a manifestation of the large dressing of the TLS energy levels by the strong coherent driving, analogous to the AC Stark shift in quantum optics that produces the Mollow triplet~\cite{Mollow1969}. In general, the Mollow triplet appears when the energy gap between the dressed states of the TLS becomes larger than their linewidth. For the parameters of \figref{fig:dissip-sq-rate}(a) and \figref{fig:squeezing-rate}(a), this condition reads $\vert\Ob\vert>\kappa_t/2$ and the side emission peaks arise at frequencies $\sqrt{\vert\Ob\vert^2-(\kappa_t/2)^2}$, indicated by the vertical dashed lines. 

The dependence of the rates $\diffrate$ and $ \sqrate$ on the TLS driving rate $\Ob$ is shown in \figref{fig:dissip-sq-rate}(b) and \figref{fig:squeezing-rate}(b) respectively, for two values of the system detuning $\dsingle$. While the coherent rate $\sqrate$ vanishes in the strong driving limit, where $\sqrate\sim s^{-1}$, the dissipative rate $\diffrate$ saturates to a value given by
\be\label{limit-diffrate}
\begin{split}
	\lim_{s \rightarrow \infty}\diffrate &= \frac{N \coup^2 }{2(\kt -\im \dsingle)}. 
\end{split}
\ee 
Note that \eqnref{limit-diffrate} is valid for any finite TLS detuning $\db$. Both the saturation of $\diffrate$ and the appearance of a Mollow triplet are identifying characteristics of incoherent resonance fluorescence spectra~\cite{CohenTannoudji1992Book,MeystreBook}. We can thus understand the rates $\sqrate$ and $\diffrate$ as stemming from the TLS incoherently pumping energy from the driving into the system or vice versa. Note that, since in our case the system does not show a continuous energy spectrum but a single resonance at frequency $\wsingle$ (in the rotating frame, $\dsingle$), it resembles more closely the more involved situation of TLS resonance fluorescence inside an optical cavity~\cite{Holm1985,Konthasinghe2012,Grunwald2013,Nguyen2011,Quang1994,Freedhoff1993,Haroche1972,Mollow1972,CohenTannoudji1977,SanchezMunozOptica2018}. Specifically, the system acts as a ``frequency filter'', probing the incoherent scattering spectrum within a narrow frequency window~\cite{Holm1985}. This is evidenced by \figref{fig:dissip-sq-rate}(b) and \figref{fig:squeezing-rate}(b) where, for $\dsingle=0$ (purple curves) only the energy scattered at the TLS natural frequency (i.e., only the central peak of the Mollow triplet) contributes to the rates, which thus monotonically depend on $\Ob$. Conversely, when the system is detuned, $\dsingle\ne 0$ (orange curves), the rates $\vert\sqrate\vert$ and $\vert\diffrate\vert$ reach a maximum at the value $\Ob$ at which the Mollow side peak and the system become resonant (compare with panel (a) of the corresponding figures).

\subsubsection{Decay rate \texorpdfstring{$\gamma$}{} and frequency shift \texorpdfstring{$\delta$}{}}
\label{sec:decay-rate}

\begin{figure}
	\centering
	\includegraphics[width=\linewidth]{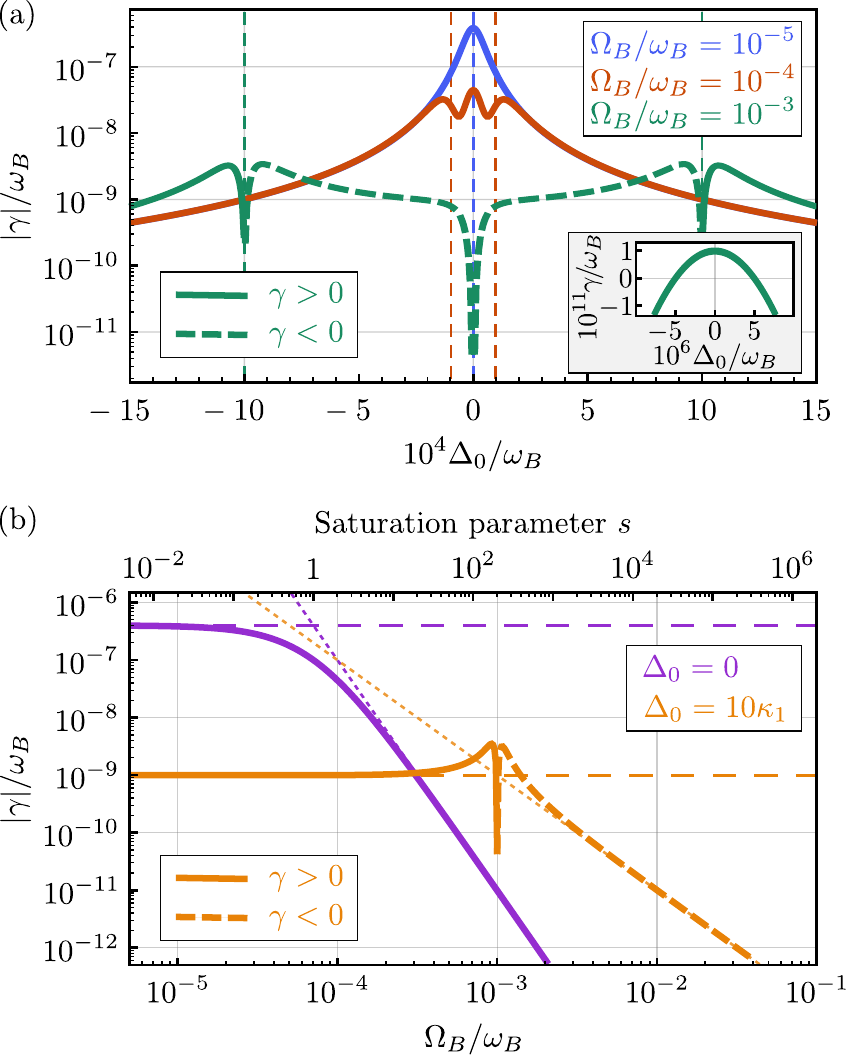}
	\caption{
		(a) Dependence of the decay rate $\gamma$ on the system detuning $\dsingle$ for $\db=0$, different TLS driving rates $\Ob$, and the parameters in Table~\ref{tab:numbers}. The dashed lines denote the frequencies of the Mollow sidebands (see main text). 
		The inset shows a close-up of the green curve at small system detunings, $\vert\dsingle\vert\lesssim\kt$.
		(b) Dependence of the decay rate $\gamma$ on TLS driving rate $\Ob$ for different system detunings $\dsingle$ and the same parameters as in panel (a). The horizontal dashed lines indicate the no-driving limit \eqnref{limit-decay-rate}, whereas the dotted lines indicate the high-driving limits \eqnref{gammalimits}.
		}
	\label{fig:decay-rate}
\end{figure}

\begin{figure}
	\centering
	\includegraphics[width=\linewidth]{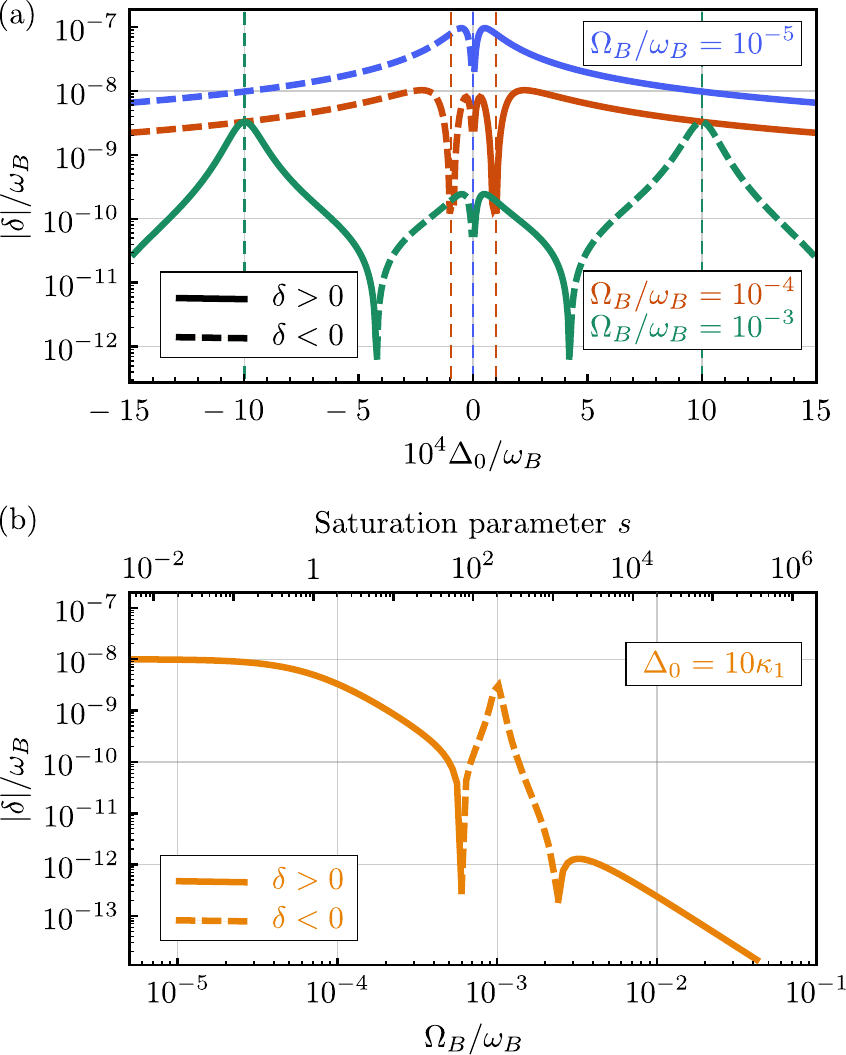}
	\caption{
	(a) Dependence of the frequency shift $\delta$ on the system detuning $\dsingle$ for $\db=0$, different TLS driving rates $\Ob$, and the parameters in Table~\ref{tab:numbers}. The dashed lines denote the frequencies of the Mollow sidebands (see main text). 
	(b) Dependence of the frequency shift $\delta$ on TLS driving rate $\Ob$ for a system detuning $\dsingle=10\kt$ and the same parameters as in panel (a).}
	\label{fig:frequency-shift}
\end{figure}

We finally focus on the decay rate $\gamma$ and the frequency shift $\delta$. Since these result from the TLS correlators
$\mv{ \sgmt{\pm i }(t+\tau)\sgmt{\mp i }(t) }$, they do not vanish in the low TLS driving limit, where 
\begin{multline}\label{limit-decay-rate}
    \lim_{s \rightarrow 0}
    \left(\begin{array}{c}
        \gamma  \\
        \delta 
    \end{array}\right)
    =
    \frac{ N \vert\coup\vert^2}{\kt^2+(\wsingle-\wb)^2} 
 	\\
 	\times \tanh\left[\frac{\hbar \wb}{2 k_B T}\right]
 	\left(\begin{array}{c}
    2\kt  \\
    \wsingle-\wb 
    \end{array}\right).
\end{multline}
The above expression for $\gamma$, which is also valid for $\Delta_B\ne 0$, coincides with the predictions of the standard tunneling model for a bath of undriven TLS~\cite{vanVleck1964}. In the opposite limit of strongly driven TLS the decay rate and the frequency shift vanish, $\lim_{s\rightarrow\infty}\gamma=\lim_{s\rightarrow\infty}\delta = 0$, as the TLS become saturated and thus induce neither absorption nor decay. The suppression of $\gamma$ for a saturated TLS bath has been demonstrated in acoustic and magnonic setups~\cite{Pfirrmann2019,Andersson2021,Heidler2021}. Both rates $\gamma$ and $\delta$ capture two different physical phenomena affecting the system: on the one hand, and similarly to the rates $\sqrate$ and $\diffrate$ analyzed in the previous section, the rates $\gamma$ and $\delta$ represent a part of the incoherent scattering of the TLS driving into the system, as they originate from the fluctuations of the TLS operators. On the other hand, they describe the contact, mediated by the TLS, between the system and the thermal reservoir inducing the TLS dissipation, as evidenced by the non-zero value of \eqnref{limit-decay-rate}. Because of the  competition between these two processes, the rates $\gamma$ and $\delta$ display a particularly rich phenomenology. 

Let us focus on the decay rate $\gamma$. This rate is shown in \figref{fig:decay-rate}(a) as a function of the system detuning $\dsingle$, for three values of the driving rate $\Ob$, and the parameters in Table~\ref{tab:numbers}. At weak driving ($s\ll1$, blue curve) it displays the usual Lorentzian profile given by \eqnref{limit-decay-rate}, whereas at higher driving (red and green curves) different regimes appear depending on the detuning $\dsingle$. In the strongly driven case $s\gg1$ the rate $\gamma$ can, remarkably, become negative indicating that, instead of damping, the TLS bath induces amplification of the system dynamics~\footnote{Note that for $\gamma<0$ the system can still be damped if its additional damping mechanisms dominate, i.e. if $\gamma_0 + \gamma>0$.}. As evidenced by \figref{fig:decay-rate}(a) and inset, amplification occurs at system-TLS detunings fulfilling $\kappa_t \lesssim\vert\dsingle\vert\le\Ob$. This behavior is captured analytically through the following limits:
\begin{multline}\label{gammalimits}
    \lim_{s\to\infty} \gamma = N\vert\coup\vert^2\kappa_1 
    \\
    \times\left\lbrace
    \begin{array}{cc}
        \dsingle^{-2} &\text{ for }\dsingle\gg\Ob\gg\kt  \\
        -\Ob^{-2} &\text{ for }\Ob\gg\dsingle\gg\kt \\
        2\kappa_1\kt \Ob^{-4} \coth\spare{\frac{\hbar\wb}{2 k_B T} } &\text{ for }\Ob\gg\kt\gg\dsingle,
    \end{array}\right.
\end{multline}
denoted by the dotted lines in \figref{fig:decay-rate}(b), where we show the dependence of $\gamma$ on the TLS driving rate $\Ob$ (solid curves). The behavior of $\gamma$, and especially the amplification $\gamma<0$, is a consequence of the non-thermal state of the TLS bath~\cite{Mollow1972}. Indeed, the TLS bath is simultaneously coupled to a thermal reservoir at temperature $T$, which tends to drive it into a thermal state, and to an external driving which, in the limit $s\gg1$, tends to drive it into a fully unpolarized state $\hat{\rho}_{B}=\mathds{1}_{2N}/2$, effectively acting as an infinite temperature thermal reservoir. For some parameter combinations and in the regime of strong driving $\Ob \gg \kt$, this results in continuous pumping of energy into the system. The behavior of the resulting amplification, captured by \figref{fig:decay-rate}, is consistent with the well-known amplification of resonance fluorescence for TLS inside an electromagnetic cavity~\cite{Wu1977,Holm1985,Haroche1972,Mollow1972,CohenTannoudji1977}. In the platforms we aim at describing in this work, namely for baths of solid-state TLS impurities affecting e.g. microwave, acoustic, or magnonic resonators, this amplification has, to our knowledge, not been reported. 

We can heuristically understand the qualitative behavior of $\gamma$ through the simpler model of a system with frequency $\dsingle$ coupled to a single, lossy TLS via Jaynes-Cummings interaction $\sim g_0(\hat{s}\hat{\sigma}_+ +\text{H.c.)}$, described by an arbitrary coupling rate $g_0$.  At resonant ($\db=0$) and strong ($s\gg1$) driving the TLS is fully dressed, i.e. it is described by the dressed eigenstates $\vert \pm \rangle = (\ket{e}\pm\ket{g})/\sqrt{2}$, where $\ket{g}$ ($\ket{e}$) denotes the TLS ground (excited) state. The energies of the dressed eigenstates are $\pm\Ob/2$. We can now write the system-TLS interaction in terms of the dressed transition matrices $\hat{\sigma}'_{\alpha\beta} = \ket{\alpha}\bra{\beta}$ ($\alpha,\beta=\pm$), and retain only the slowly oscillating terms under a rotating wave approximation. The validity of the rotating wave approximation and the form of the resulting interaction Hamiltonian depend on $\dsingle$. Specifically, for $\kt \ll \vert\dsingle\vert \sim \Ob$ and assuming $\vert\dsingle-\Ob\vert,g_0\ll \vert\dsingle\vert + \Ob$, the Hamiltonian reads $\sim ( \hat{s}\hat{\sigma}'_{+-}+\text{H.c.})$, while for $\vert\dsingle\vert\lesssim\kt\ll \Ob$ and assuming $g_0\ll \Ob$ it reads $\sim(\hat{s}+\hat{s}^\dagger)(\hat{\sigma}_{++}'-\hat{\sigma}'_{--})$. Let us examine in these cases the system emission and absorption rates, $\gammaC$ and $\gammaB$, and the total decay rate $\gamma=\gammaC-\gammaB$. We further simplify our toy model by assuming only energy-conserving, first-order processes contribute to these rates. First, in the case $\kt \ll \vert\dsingle\vert$, system emission and absorption involve a single transition between the two TLS dressed states plus, if $\dsingle\ne\Ob$, an additional energy exchange between the TLS and its own thermal reservoir in order to conserve energy. By examination of the respective processes one can infer that $\gammaC\to\gammaB$ if $\dsingle=\Ob$, and  $\gamma_\alpha\propto\bar{n}(\vert\Ob-\vert\dsingle\vert\vert) + \delta_{\alpha,\text{sign}[\Ob-\vert\dsingle\vert]}$ if $\dsingle\ne\Ob$, with $\delta_{\alpha,\beta}$ the Kronecker delta. This simple model thus qualitatively captures the positive (negative) values of $\gamma$ for positive (negative) values of $\Ob-\vert\dsingle\vert$ (see \figref{fig:decay-rate}(a)). In the second case, namely the regime $\vert\dsingle\vert\lesssim\kt$, the interaction contains only the matrices $\hat{\sigma}_{\alpha\alpha}'$, and hence absorption and emission processes involve no transition between the dressed TLS states. The system thus effectively exchanges energy directly with the thermal reservoir of the TLS which, being in thermal equilibrium, necessarily results in $\gamma>0$. Our heuristic argument thus also captures the behavior of $\gamma$ at small $\dsingle$ (inset of \figref{fig:decay-rate}(a)). 

For completeness we display in \figref{fig:frequency-shift}(a) the frequency shift $\delta$ as a function of the system detuning $\dsingle$ for the same parameters as in \figref{fig:decay-rate}. In the weak driving regime (blue curve) the frequency shift displays the form given by Eq.~\eqref{gammalimits}. This is the usual profile obtained when computing the electromagnetic response function of two-level systems e.g. in atomic optics~\cite{CohenTannoudji1992Book}. At stronger driving rates, peaks emerge at the Mollow sideband frequencies, confirming the incoherent scattering contribution. The dependence of the frequency shift $\delta$ on the TLS driving rate $\Ob$ is shown in \figref{fig:frequency-shift}(b) for the same parameters as in \figref{fig:decay-rate}. For $\dsingle=0$ the frequency shift is exactly zero.

\subsection{Steady-state properties: squeezing and dynamical instabilities}
\label{sec:dynamics}

\begin{figure}[t]
\centering
\includegraphics[width=\linewidth]{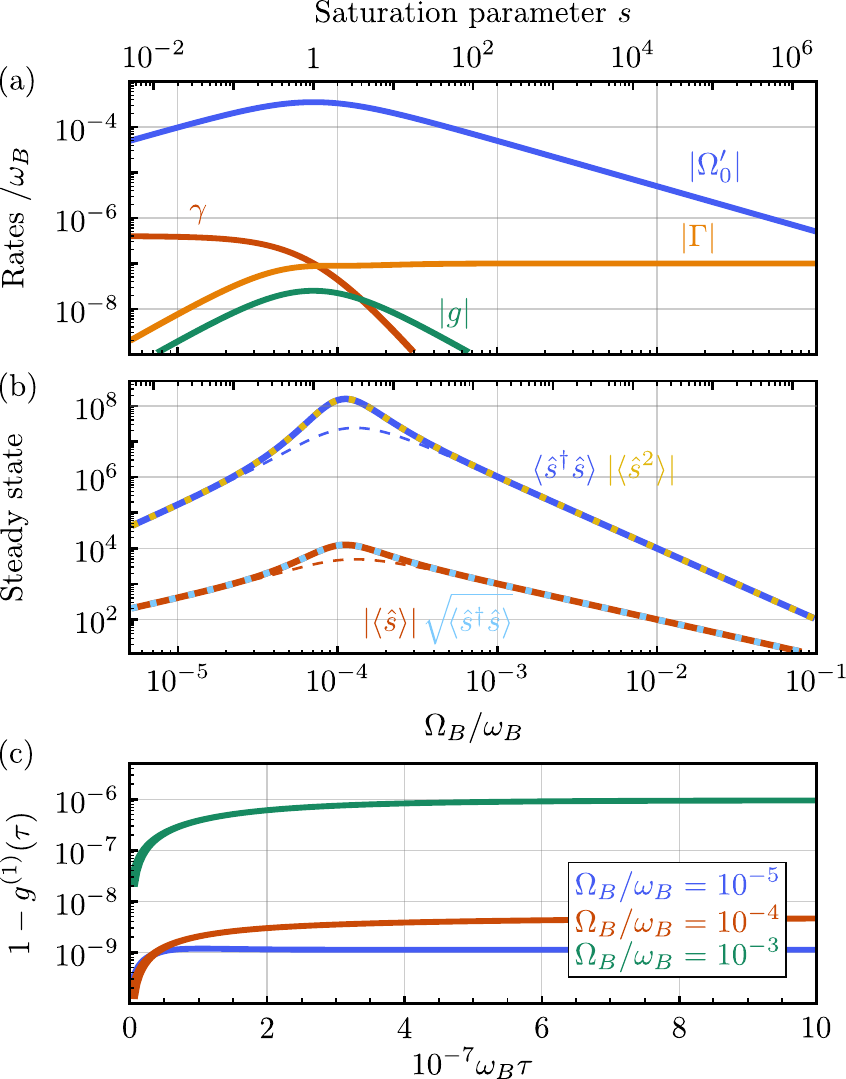}
\caption{
	(a) Master equation rates as a function of TLS driving rate $\Ob$ for the resonant case $\wsingle=\wb=\omega_d$. The frequency shift in this case is $\delta=0$.
	(b) Steady-state value of the system operators (solid and dotted curves) as a function of TLS driving rate $\Ob$. The dashed curves show the approximate solution given by \eqnref{sds-approx}. 
	(c) First-order coherence function $g^{(1)}(\tau)$, \eqnref{g1}, for three different TLS driving rates $\Ob$. In all panels we take the parameter values of Table~\ref{tab:numbers}.
    }
\label{fig:dynamics}
\end{figure}

In this final section we focus on the impact of the TLS bath on the system, specifically on its steady state. We consider a relevant particular case, namely an undriven system ($\Osingle=0$) in resonance both with the TLS and with their driving, i.e. $\wsingle=\wb=\wdr$. The master equation rates in this case are shown in \figref{fig:dynamics}(a) for the parameters of Table~\ref{tab:numbers}. For these parameters the bath does not induce amplification, as $\gamma +\gamma_0 >0 $ $\forall s$. Among the three  terms resulting from TLS correlators, namely $\gamma$, $\sqrate$, and $\diffrate$, the decay rate $\gamma$ dominates at low saturation, $s\lesssim 1$ whereas the rate $\diffrate$ dominates at high saturation, $s\gtrsim 1$. Any exotic effect ascribed to the non-thermal state of the TLS will thus appear in the regime of unconventional dissipation $s\gtrsim 1$. The squeezing rate $g$ is typically much smaller than the dissipation rates $\gamma$ and $\diffrate$. Regardless of the predominant dissipation, and as shown by \figref{fig:dynamics}(a), the effective dynamics are dominated by the coherent driving rate $\Oeff$. The strong impact of this effective driving has been experimentally observed in high-$Q$ microwave cavities~\cite{Heidler2021}. 

Since the master equation \eqnref{master-equation} is quadratic, the steady state is Gaussian~\cite{Nicacio2016}. It is completely determined by the first and second-order momenta, $\bm{v}\sts=(\mv{\sopd \sop},\mv{\sop}, \mv{\sopd},\mv{\sop^2},  \mv{(\sopd)^2})\sts^T = -A_S^{-1}\bm{a}_S$, where the matrix $A_S$ and the vector $\bm{a}_S$ are given by Eqs.~\eqref{sds-system} and \eqref{sds-independentvector}, respectively. In \figref{fig:dynamics}(b) we show the steady-state values $\vert\mv{\sop}\sts\vert$ (solid red curve), $\mv{\sopd \sop}\sts$ (solid blue curve), and $\vert\mv{\sop^2}\sts\vert$ (doted yellow curve), for the same parameters as in \figref{fig:dynamics}(a). The thin dashed red and blue curves in the figure correspond to the limit $g\to 0$, where the steady-state is analytically approximated by
\begin{equation}\label{sds-approx}
    \mv{\sopd\sop}\sts \approx \frac{\gammaB}{\gamma} + \frac{4\vert\Oeff\vert^2}{\gamma^2} \hspace{0.5cm} ; \hspace{0.5cm} \mv{\sopd}\sts \approx \frac{2i\Oeff}{\gamma}.
\end{equation}
As shown by \figref{fig:dynamics}(b) the above expressions are very similar to the exact solution except in the vicinity of $s= 1$. Thus, neglecting the small rate $\sqrate$ is a good approximation to obtain the values of $\vert\mv{\sop}\sts\vert$ and $\mv{\sopd \sop}\sts$. Since the effective coherent driving, given by the rate $\Oeff$, is the dominant effect in the master equation, the steady state of the system is close to a coherent state, i.e., $\mv{\sopd \sop}\sts \approx \vert\mv{ \sop}\sts\vert^2 \approx \vert\mv{ \sop^2}\sts\vert$, as indicated by \figref{fig:dynamics}(b). This is confirmed by \figref{fig:dynamics}(c) where we display the first-order coherence of the steady state, defined as~\cite{CohenTannoudji1992Book,Carmichael}
\be\label{g1}
    g^{(1)}(\tau) = \frac{\mv{\sopd(0)\sop(\tau)}\sts}{\mv{\sopd \sop}\sts},
\ee
for different values of the TLS driving rates corresponding to $s \ll 1$ (blue curve), $s\sim 1$ (red curve), and $s\gg 1$ (green curve). Although due to the dissipative dynamics the first-order coherence deviates from $1$, it remains close to this value at all times, confirming the quasi-coherent nature of the steady state.

\begin{figure}[t]
	\centering
	\includegraphics[width=\linewidth]{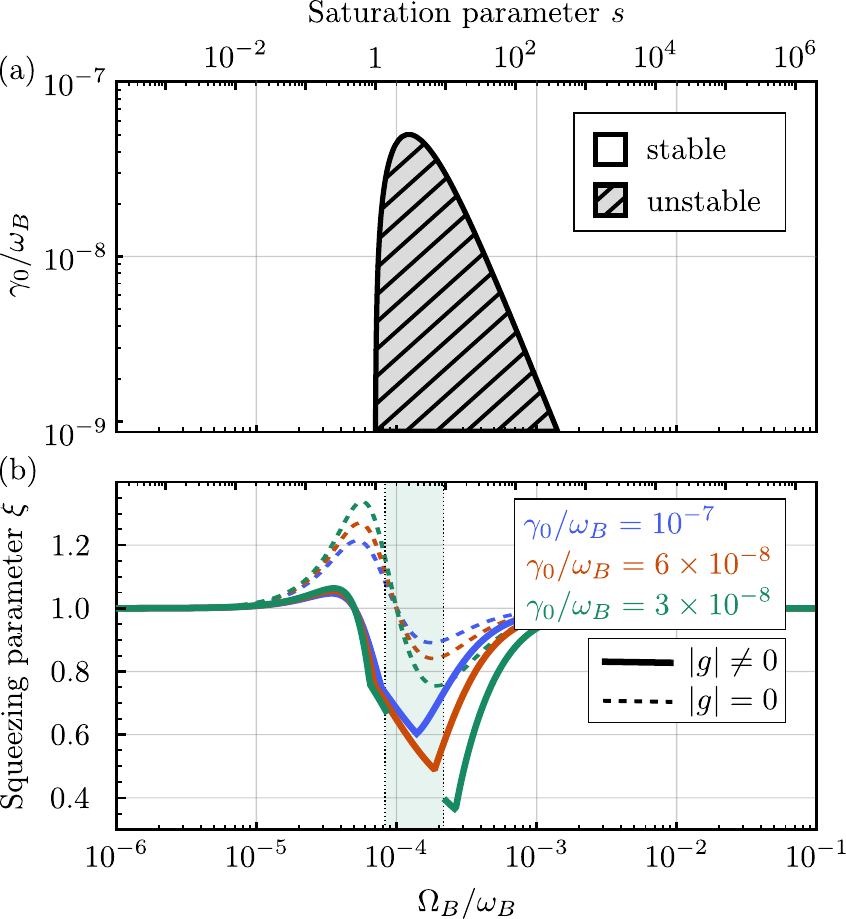}
	\caption{
	(a) System stability diagram as a function of the TLS driving rate $\Ob$ and the system decay rate $\gamma_0$, obtained by considering the stability criterion \eqnref{stability-criterion}.
	(b) Dependence of the squeezing parameter $\xi$ on TLS driving rate $\Ob$, for different system decay rates $\gamma_0$. Solid curves: full model. Dashed curves: Solution for $g\to 0$. For the smallest decay rate depicted here (green curve), the squeezing parameter $\xi$ is not well-defined in the unstable region denoted by the shaded green area. 
	}
	\label{fig:squeezing}
\end{figure}

Despite being close to a coherent state, the steady state is far from the conventional steady state of a driven lossy harmonic oscillator. To show this, we first focus on the conditions for the existence of a steady state. The master equation \eqnref{master-equation} has a steady state if and only if the system of equations governing $\bm{v}\sts$ is linearly stable~\cite{Nicacio2016}, i.e. if $\max[\text{Re}(\lambda_j)]\le 0$, with $\lambda_j$ ($j=1,...,5$) the eigenvalues of $A_S$. In the resonant case under study, $\wsingle=\wb=\wdr$, the system is linearly stable if
\be \label{stability-criterion}
	\gamma_0+\gamma \geq 4\vert g\vert.
\ee
In \figref{fig:squeezing}(a) we show a stability diagram for the system, as a function of the bare system linewidth $\gamma_0$ and the TLS driving rate $\Ob$. For a saturated bath ($s\gg 1$) or for a bath in thermal equilibrium ($s\ll1$) the system is always stable as $\sqrate \to 0$ and $\gamma$ either vanishes or is always positive. Stability is also guaranteed at large enough $\gamma_0$, where the decay of the system to its additional bath (assumed in equilibrium, see \secref{sec:model-derivation}) dominates over the TLS-induced dissipation. However, at intermediate values of $s$ the system can become dynamically unstable. Remarkably, the dynamical instability can originate either from a strong amplification ($\gamma<0$, see previous section) or, as is the case in \figref{fig:squeezing}(a), from a large enough value of the squeezing rate $\sqrate$~\cite{Kustura2019}. 
Note that since the small rate $\sqrate$ plays a relevant role in determining stability, the approximations at $g\to 0$ (\eqnref{sds-approx}) become inaccurate when the system is near the instability regime. The dynamics in this critical, near-unstable regime are very sensitive to the value of $\sqrate$, a property that could be used to accurately measure this rate.

The second unconventional feature of the system steady state is the presence of squeezing. We analyze squeezing in the stable regime, where the steady state exists and it is Gaussian, by computing its covariance matrix~\cite{Simon1994}
\be \label{cov-matrix}
\begin{split}
	\sigma=
	\begin{pmatrix}
		V_x    & C_{xp}   \\
		C_{xp} & V_p    
	\end{pmatrix},
\end{split}
\ee
where $V_x=\mv{(\xop-\mv{\xop})^2}$, $V_p= \mv{(\popr-\mv{\popr})^2}$, and $C_{xp}=\lbrace \xop-\mv{\xop},\popr-\mv{\popr}\rbrace/2$; with $\xop = (\sop+\sopd)/\sqrt 2$ and $\popr = \im(\sopd-\sop)/\sqrt{2}$ being the quadrature operators. The system squeezing can be quantified via the squeezing parameter $\xi = 1/\sqrt{2 \text{min}_k (\lambda_k)}$, where $\lambda_k \in \mathbb{R}^+$ are the eigenvalues of \eqnref{cov-matrix}~\cite{Simon1994}. The state is squeezed when $\xi>1$, and larger values of $\xi$ correspond to larger squeezing. In \figref{fig:squeezing}(b), we plot the squeezing parameter $\xi$ (solid curves) as a function of the TLS driving rate for different values of the system decay rates $\gamma_0$. The shaded green area marks the instability window for $\gamma_0/\wb = 3\times10^{-8}$ (compare with \figref{fig:squeezing}(a)). According to \figref{fig:squeezing}(b) the steady state of the system is squeezed for a range of saturation parameters around $s\approx0.1$. Larger squeezing is attained at low values of $\gamma_0$, where the TLS-induced dissipation dominates over the system intrinsic dissipation. The steady-state squeezing has both coherent and dissipative contributions, coming from the rates $\sqrate$ and $\diffrate$, respectively. This is proven by the dashed curves in \figref{fig:squeezing}(b), which depict the squeezing parameter obtained under the substitution $g\to 0$. Remarkably, the squeezing is reduced in the presence of both mechanisms, i.e. when $\sqrate,\diffrate\ne0$, as the corresponding terms in the master equation induce squeezing in mutually orthogonal directions in phase space. The comparison between the dashed and solid curves in \figref{fig:squeezing}(b) also shows that, although approximating $g\to0$ remains a useful approximation for some steady-state properties such as the occupation number $\mv{\sopd\sop}\sts$, it critically fails to capture others such as stability and squeezing. We finally note that the steady-state squeezing is in principle experimentally observable in magnonics or acoustic platforms, where the regime $s\approx1$ can be achieved~\cite{Behunin2017,Andersson2021,Pfirrmann2019,Kosen2019}.

\section{Conclusion}
\label{sec:conclusion}

We have developed a theoretical model describing the effective dynamics of a system in the presence of a coherently driven two-level-system (TLS) bath. This has been done by explicitly tracing out the bath degrees of freedom to obtain a Born-Markov master equation. In the limit of weak TLS driving, our results recover the expression given by the standard tunneling model for undriven TLS baths. In the limit of strong TLS driving, our model predicts a vanishing linewidth due to saturated TLS, as observed in experiments. In the intermediate driving regime exotic dynamics arise as the state of the TLS bath is maximally out of thermal equilibrium. Specifically, the TLS can induce linear instability of the system, either through negative linewidth (amplification) or through single-mode squeezing. Moreover, in the linearly stable regime the steady state of the system is squeezed. To our knowledge, most of these predictions have not been experimentally observed. From a theoretical point of view, an interesting outlook consists of characterizing the non-Markovian effects arising for non-monochromatic TLS driving~\cite{Reichert2016,Grabert2018,Grabert2016}, or for more strongly coupled and/or less lossy TLS baths.

To conclude, our model provides a theoretical tool for studying, from the quantum optics perspective, the complex TLS baths affecting most quantum technological platforms. It proves that external driving of these baths can be used as a tool not only to minimize dissipation, e.g. by saturating the TLS, but also to probe the TLS bath and acquire deeper information about its properties. 

\begin{acknowledgments}
We thank A. Gonzalez-Tudela and Y.~Nakamura for helpful discussions. 
C. G.-B. acknowledges support from the European Union (PWAQUTEC, H2020-MSCA-IF-2017, no. 796725).
\end{acknowledgments}

\bibliography{DrivenTLSBibliography}
\end{document}